\documentclass[sigconf,natbib=true,anonymous=false]{acmart}

\makeatletter
\def\@ACM@checkaffil{
    \if@ACM@instpresent\else
    \ClassWarningNoLine{\@classname}{No institution present for an affiliation}%
    \fi
    \if@ACM@citypresent\else
    \ClassWarningNoLine{\@classname}{No city present for an affiliation}%
    \fi
    \if@ACM@countrypresent\else
        \ClassWarningNoLine{\@classname}{No country present for an affiliation}%
    \fi
}
\makeatother

\usepackage{natbib}
\setcitestyle{square, numbers, sort}
\usepackage{graphicx}
\usepackage{tabularray}

\AtBeginDocument{%
  }

\setcopyright{acmcopyright}
\copyrightyear{2023}
\acmYear{2023}
\acmDOI{XXXXXXX.XXXXXXX}

\acmConference[KDDMLF 2023]{Machine Learning in Finance 2023@KDD}{Aug 07,
  2023}{Long Beach, CA}
\acmPrice{15.00}
\acmISBN{978-X-XXXX-XXXX-X/YY/MM}

\begin{document}

\title{BIRP: Bitcoin Information Retrieval Prediction Model Based on Multimodal Pattern Matching}

\author{Minsuk Kim}
\authornote{These authors contributed equally to this research.}
\affiliation{%
  \institution{MINDsLab Inc.}
}
\email{minsuk@mindslab.ai}

\author{Byungchul Kim}
\authornotemark[1]
\affiliation{%
  \institution{HighDev}
}
\email{kimkevin2657@naver.com}

\author{Junyeong Yong}
\authornotemark[1]
\affiliation{%
  \institution{MINDsLab Inc.}
}
\email{yjy2026@mindslab.ai}

\author{Jeongwoo Park}
\affiliation{
    \institution{Pukyong National University}
}
\email{Russell888@naver.com}

\author{Gyeongmin Kim}
\authornotemark[1]
\authornote{Corresponding author.}
\orcid{0000-0002-2851-0374}
\affiliation{%
  \institution{Korea University}
}
\email{totoro4007@gmail.com}





\renewcommand{\shortauthors}{Minsuk et al.}

\begin{abstract}
Financial time series have historically been assumed to be a martingale process under the Random Walk hypothesis. Instead of making investment decisions using the raw prices alone, various multimodal pattern matching algorithms have been developed to help detect subtly hidden repeatable patterns within the financial market. 
Many of the chart-based pattern matching tools only retrieve similar past chart (PC) patterns given the current chart (CC) pattern, and leaves the entire interpretive and predictive analysis, thus ultimately the final investment decision, to the investors. 
In this paper, we propose an approach of ranking similar PC movements given the CC information and show that exploiting this as additional features improves the directional prediction capacity of our model. We apply our ranking and directional prediction modeling methodologies on Bitcoin due to its highly volatile prices that make it challenging to predict its future movements.
\end{abstract}




\maketitle

\section{Introduction}
Since the introduction of Bitcoin (BTC) in 2008 \cite{nakamoto2008bitcoin}, the cryptocurrency (crypto) market has grown significantly.
As a result, it attracted many investors and researchers attempting to forecast the movement of these crypto assets in search of profits.

Technical analysis is a discipline that analyzes the statistical transformation of the undelying financial price and volume time series. In this paper, we apply Balance of Power (BOP), Even Better Sinewave (EBSW), Chaikin Money Flow (CMF), Differencing (DIFF), and Inter Ratios (INTRA), which is technical analysis on the historical Bitcoin prices and volume to help rank past chart (PC) movements given the current chart (CC) movements. In this work, we attempt to compare predictive capabilities of various ranking methodologies between PC patterns given CC patterns. 
Our contribution is mainly two folds:
\begin{itemize} 
\item We propose four different ranking methodologies for chart pattern matching and rank similar PC segments based on proposed metrics. 
\item We propose a BTC directional forecasting model for trading BTCUSDT perpetual and show that using voting information from pattern matched chart segments in the past improve the performance of our forecasting model.
\end{itemize}  

\section{Related work}
There are numerous researches on how to detect chart patterns such as \cite{gong2013comparison} and \cite{velay2018stock}. However, it is challenging to find literature that uses detected chart patterns for further modeling or using such information to devise a trading strategy.
There are previous approaches that use similar patterns for modeling but they do not apply these techniques specifically in the domain of finance, much less crypto or BTC\cite{velay2018stock, avva2015pattern}.
Moreover, there are chart pattern detecting applications available for the traders such as those provided by TrendSpider \footnote{http://trendspider.com/} and BTC pattern calculator \footnote{https://miningcalc.kr/chart/btc}, but they simply detect the patterns and do not take it a step further.
Our research is also closely related to CBITS \cite{kim2023cbits} as we incorporate our Information Retrieval (IR) based feature engineering (FE) technique into the CBITS framework. Furthermore, we use one of the crypto language models (LM) introduced in CBITS, crypto DeBERTa, a transformer based LM that improves upon BERT \cite{devlin-etal-2019-bert} by exploiting disentangled attention and enhanced mask decoder \cite{he2021deberta} for our multimodal embedding based ranking method.  


\section{Problem Definition}
In this section, we define the problem for the proposed methods of our model. We approach BTC price directional forecasting task as a three class classification problem. The labels are defined as follows:

\[ u_{t+1} = \frac{\text{high}_{t+1} - \text{close}_t}{\text{close}_t}, v_{t+1} = \frac{\text{low}_{t+1} - \text{close}_t}{\text{close}_t} \]

\begin{itemize} 
\item $c_0:$ BTC price rises by at least 0.75\% within the next 4 hours i.e. $u_{t+1} >= 0.0075$. 
\item $c_1:$ BTC price drops by at least 0.75\% within the next 4 hours i.e. $v_{t+1} <= -0.0075$.
\item $c_2:$ BTC price change within the next 4 hours is less than 0.75\% i.e $u_{t+1} < 0.0075$ and $v_{t+1} > -0.0075$.  
\end{itemize} 

The label $c_0$ translates to long position, the label $c_1$ translates to short position and $c_2$ translates to holding (taking no action). If both $u_{t+1} >= 0.0075$ and $v_{t+1} <= -0.0075$ occur for the next 4 hours, then we gave the labeling priority to $c_0$ (i.e. when both taking long or short results in at least 0.75\% profit then we simply label that timestep as long). 

\subsection{Dataset Description} 
We collected 4 hourly BTC/USDT data from Binance \footnote{www.binance.com/en}, one of the largest crypto exchanges, and we ended up with 11,812 data points ranging from 2017-08-23 16:00:00 to 2023-01-15 20:00:00.
After labeling the data, we end up with a label distribution of approximately 48.11\% for $c_0$, 29.40\% for $c_1$, and 22.49\% for $c_2$.
Out of the 11,812 data points, we use the first 80\% of the data as candidates (9,449 data points) for pattern matching and the rest for experimentation. 
2,363 data points ranging from 2021-12-18 04:00 to 2023-01-15 20:00:00 were split into train/validation/test dataset in 8:1:1 ratio, and each of these 2,363 data points were compared with the candidates to find its similar counterparts in the past.
For each of the 2,363 data points, we collected the top 30 similar chart patterns for each of the four different similarity calculation strategies.

\subsection{Modeling}

\begin{figure}[h]
    \centering
    \includegraphics[width=0.4\textwidth]{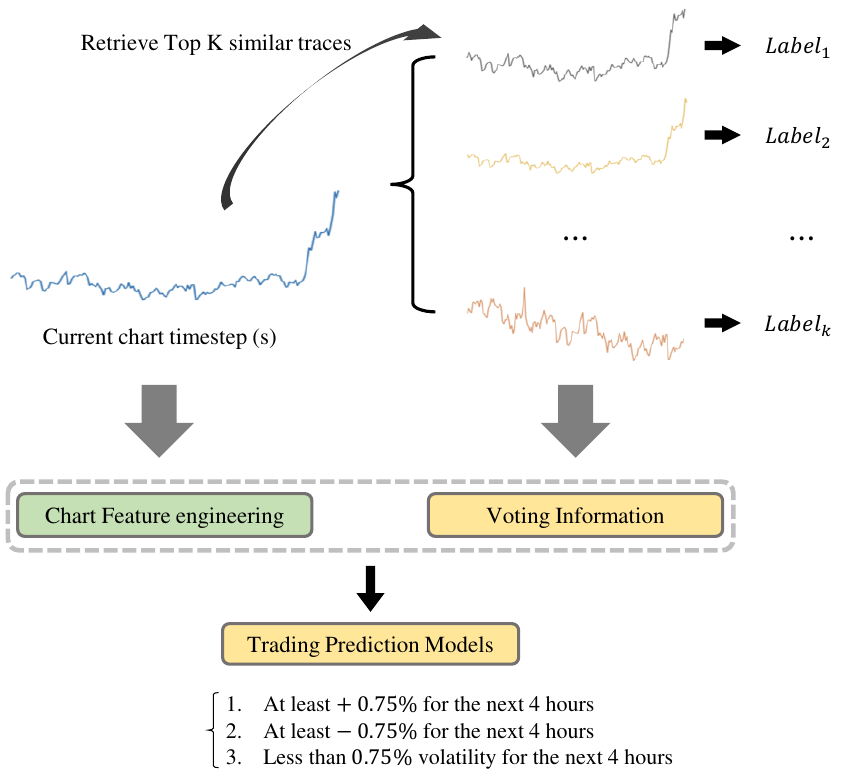}
    \caption{IR-based FE System for modeling BTC directional prediction.}
    \label{figure:workflow}
\end{figure}

We employ XGBoost\cite{xgboostpaper} as our directional forecasting model as it is fast to train and is robust for tabular data based classification tasks. 
To highlight some important hyper-parameters, we used 200 for the number of boosting rounds with a learning rate of 0.3. The maximum tree depth for base learners was set to 6, and the tree method was set to "gpu\_hist". We also considered the class weights of the train dataset when training XGBoost. 
Essentially we compare the performance of when we use similar past chart information or not. We will denote these two cases as a system with IR-based FE and a system without IR-based FE. The overall approach is illustrated in Figure \ref{figure:workflow}. We compare the performances of each method by calculating accuracy and weighted F1 score.

\subsubsection{Without IR-based FE} \label{sec:Without IR-based FE}
When we do not use IR-based FE we simply use chart based features only as inputs to XGBoost for training. Most of the chart features that were used are features focused on calculating volatility or simply ratios of the open, high, low, close and volume features. Although XGBoost does not require feature scaling/normalization, in order to make the training more stable we purposely chose features with similar value ranges. 

\begin{itemize}
\item \textbf{BOP}: The balance of power (BOP) is an oscillator that measures the strength of the buy and sell pressures. When BOP is positive it suggests that the market is bullish and vice-versa. BOP that is close to zero indicates a balance between the two powers and it may signify a trend reversal. 
\item \textbf{EBSW}: The even better sinewave (EBSW) is a variation of the Hilbert sine wave, and it is an indicator that can inform the model about the bullish and bearish cycle of prices. 
\item \textbf{CMF}: The chaikin money flow (CMF) is an indicator used to monitor both the accumulation and distribution of an asset over a specified period. The default period of 20 is used for CMF.  
\item \textbf{DIFF}: This is identical to the differencing features presented in \cite{kim2023cbits}. It is the ratio of raw chart features across different periods. The first differencing of close prices would be calculated as 
\[
\text{First Difference of Close} = \frac{\text{close}_t}{\text{close}_{t-1}}
\] 
and in general the Kth differencing of the close prices is simply 
\[
\text{Kth Difference of Close} = \frac{\text{close}_t}{\text{close}_{t-K}}
\]
Similarly we carry out this differencing procedure for all open, high, low, close and volume and used K = 1,2,...,12 for differencing.  
\item \textbf{INTRA}: This feature is a ratio of different features in the current timestep. Specifically we used $\frac{\text{high}_t}{\text{low}_t}$, $\frac{\text{high}_t}{\text{open}_t}$, $\frac{\text{low}_t}{\text{open}_t}$, $\frac{\text{close}_t}{\text{open}_t}$, $\frac{\text{high}_t}{\text{close}_t}$, $\frac{\text{low}_t}{\text{close}_t}$. 
\end{itemize} 
after pre-processing the chart features we end up with a total of 68 features for training. 

\subsubsection{With IR-based FE}
We use the same chart features presented in section \ref{sec:Without IR-based FE} in addition to the vote information from past similar ranked chart features. 
Given a timestep $t$, we retrieve at most top 30 timesteps $t'$ that is the most similar to the timestep $t$ under one of the four ranking methods. 
Then we separately calculate the performance of the model's directional forecast when we use voting information from the top $K \in \{5, 10, 15, 20, 25, 30\}$ most similar past patterns. Given top $K$ similar $t'$, we count the number of cases when the labels were $c_0, c_1$ or $c_2$. 
For example, if $K=10$ and out of those top $K$ similar instances if 6 of them were $c_0$, 3 were $c_1$, and 1 was $c_2$, then the voting vector can be formed as $(6, 3, 1)$. 
Before we use this as additional input features to the XGBoost model we softmax normalize these scores.
For example, the count for $c_0$ will be normalized as follows: 
\[
\text{cnt($c_0$)} \rightarrow \frac{e^{\text{cnt($c_0$)}}}{e^{\text{cnt($c_0$)}} + e^{\text{cnt($c_1$)}} + e^{\text{cnt($c_2$)}}} 
\]

Similar transformation is applied to the count for $c_1$ and $c_2$. After repeating this procedure for all $K$, we calculate the average accuracy and weighted F1 scores for final comparison with the model that does not use IR-based FE.

\section{Ranking Methods}
In this section we describe our ranking approaches in detail. For all the ranking methods we ranked the top 30 most similar past timesteps given the query timestep.  

\subsection{Euclidean Distance} 
Our data is represented as follows: for timestep $t$.
\[
p_t = [f_{t,1}, f_{t,2}, ..., f_{t, m}] 
\]
where $f$ denotes a feature and $m$ is the total number of features (in our case 68 as explained in section 3.2.1). Given some timestep $t' < t$ the euclidean distance (L2 norm) is calculated via the following formula:
\[
SIM(p_t, p_{t'}) = \sqrt{(f_{t,1}-f_{t',1})^2 + ... + (f_{t, m} - f_{t', m})^2}
\]
we retrieve top 30 such $p_t'$ that has the smallest euclidean distance to $p_t$. 

\subsection{TS2Vec} 
The timeseries to vector (TS2Vec) embedding method was first proposed in \cite{ts2vecpaper} and it has proved to effectively extract time series representations that are task agnostic. There are two major components to the TS2Vec architecture: 
\begin{itemize}
    \item The TS2Vec encoder consists of the input projection layer, the timestamp masking layer and the dilated convolutions layer in this order. The timestamp masking layer randomly binary masks the latent vector and this idea was motivated by \cite{simclrpaper} and \cite{simcsepaper} to create augmented context views. The encoder is then optimized via the temporal contrast loss and instance-wise contrastive loss. 
    \item The input time series is randomly cropped into two different time series with overlapping timesteps for positive pair creation in an unsupervised setting. 
\end{itemize}

Due to its design the TS2Vec requires a time series of length $l > 1$ and we set $l = 6$, or 24 hours worth of time frame since we are dealing with 4 hourly chart data. We first train the TS2Vec encoder for 100 epochs, batch size 16, learning rate 0.001, hidden dimension size 64 and output dimension size 128 on NVIDIA A100-80GB GPU. The TS2Vec encoder is trained only on the candidate pool and not on the train/validation/test dataset. Afterwards, we calculate the TS2Vec embeddings for the train/validation/test dataset and all the embeddings for the candidate pool, and calculate the cosine distance between the embeddings. We obtain the top 30 embeddings based on the closeness of the cosine distances. Given that the query is $X_t=[x_{t-5}, ..., x_{t}]$ and the candidate is $X_{t'}=[x_{t'-5}, ..., x_{t'}]$:
\[
Q_{\text{chart\_emb}} = \text{TS2Vec}(X_t),  C_{\text{chart\_emb}} = \text{TS2Vec}(X_{t'}) 
\]

\[
\text{SIM}(Q_{\text{chart\_emb}}, C_{\text{chart\_emb)}}) = 1 - \frac{Q_{\text{chart\_emb}} \cdot C_{\text{chart\_emb}}}{||Q_{\text{chart\_emb}}|| ||C_{\text{chart\_emb}}||}
\]

\begin{table*}
\centering
\footnotesize
\begin{tblr}{
  cells = {c},
  cell{2}{1} = {c=9}{},
  cell{5}{1} = {c=9}{},
  cell{8}{1} = {c=9}{},
  cell{11}{1} = {c=9}{},
  vline{2} = {3-4,6-7,9-10,12-13}{},
  hline{1-3,5-6,8-9,11-12,14} = {-}{},
}
                                                           & \textbf{No FE} & \textbf{Top 5} & \textbf{Top 10} & \textbf{Top 15} & \textbf{Top 20~} & \textbf{Top 25} & \textbf{Top 30} & \textbf{Average}        \\
\textbf{Ranking Strategy 1. Random Sampling}               &                &                &                 &                 &                  &                 &                 &                         \\
Accuracy(\%)                                               & 51.899         & \underline{54.084(+2.185)} & 53.873(+1.974)  & 53.608(+1.709)  & 53.485(+1.586)   & 53.840(+1.941)  & 53.658(+1.759)  & \textbf{53.758(+1.859)} \\
F1 score                                                   & 0.580          & \underline{0.598(+0.018)}  & 0.595(+0.015)   & 0.593(+0.013)   & 0.592(+0.012)    & 0.595(+0.015)   & 0.593(+0.013)   & \textbf{0.595(+0.015)}  \\
\textbf{Ranking Strategy 2. Euclidean Distance}            &                &                &                 &                 &                  &                 &                 &                         \\
Accuracy(\%)                                               & 51.899         & 51.477(-0.422) & 55.274(+3.375)  & 56.118(+4.219)  & 52.321(+0.422)   & \underline{56.540(+4.641)}  & 56.118(+4.219)  & \textbf{54.641(+2.742)} \\
F1 score                                                   & 0.580          & 0.579(-0.001)  & 0.611(+0.031)   & 0.614(+0.034)   & 0.583(+0.003)    & \underline{0.622(+0.042)}   & 0.616(+0.036)   & \textbf{0.604(+0.024)}  \\
\textbf{\textbf{Ranking Strategy 3. TS2Vec Embedding}}     &                &                &                 &                 &                  &                 &                 &                         \\
Accuracy(\%)                                               & 51.899         & 56.540(+4.641) & \underline{58.650(+6.751)}  & 50.211(-1.688)  & 53.586(+1.687)   & 51.899(+0.000)  & 52.743(+0.844)  & \textbf{53.938(+2.039)} \\
F1 score                                                   & 0.580          & 0.616(+0.036)  & \underline{0.635(+0.055)}   & 0.563(-0.017)   & 0.593(+0.013)    & 0.578(-0.002)   & 0.582(+0.002)   & \textbf{0.595(+0.015)}  \\
\textbf{\textbf{Ranking Strategy 4. Multimodal Embedding}} &                &                &                 &                 &                  &                 &                 &                         \\
Accuracy(\%)                                               & 51.899         & \underline{57.806(+5.907)} & 51.899(+0.000)  & 55.274(+3.375)  & 56.540(+4.641)   & 57.384(+5.485)  & 55.696(+3.797)  & \textbf{55.767(+3.868)} \\
F1 score                                                   & 0.580          & \underline{0.628(+0.048)}  & 0.578(-0.002)   & 0.610(+0.030)   & 0.615(+0.035)    & 0.624(+0.044)   & 0.609(+0.029)   & \textbf{0.610(+0.030)}  
\end{tblr}
\label{table:experimental results}
\caption{Experiment results of the four ranking strategies. Each average score is marked with bold and underline scores indicate the highest model performance in each rows.}
\end{table*}

\subsection{Multimodal} 
The multimodal method of ranking involves the use of both chart and news data to generate the embeddings for the time series. It also uses the same length $l = 6$ and calculates the TS2Vec embedding of that time series first, then additionally computes the average news embedding between times $t-1$ and $t$ by using Crypto DeBERTa\cite{kim2023cbits}. 
If the current timestep is $t$ then we use information from timesteps $[t-5,...,t]$ to calculate the TS2Vec embedding and gather all the news that were released in $[t-1, t]$ (not the entire 24 hours but just the past 4 hours) to calculate the average news embedding in this timeframe. We simply extract the \textit{[CLS]} embedding of DeBERTa's output for each news and average these \textit{[CLS]} token representations. 
Then the average news embedding and the TS2Vec embedding are summed to create a multimodal embedding and similar to 4.2, we use the cosine distances of these embeddings to get the top 30 most similar past timesteps. Before summing the average news embedding and the TS2Vec embedding, we shrink the news embedding dimension from $768 \rightarrow 128$ by applying uniform manifold approximation and projection\cite{umappaper}. 
For the news data we use Coinness Korea\footnote{https://coinness.com/}, which is also followed in \cite{kim2023cbits}. 
Multimodal embeddings should allow us to capture both the news sentiments and chart dynamics when searching for past patterns.

\[
\bar Q_{\text{news\_emb}} = \frac{1}{N}\sum_{i=1}^N \text{LM}(\text{news}_i), Q_{\text{chart\_emb}} = \text{TS2Vec}(p_t) 
\]

\[
\bar C_{\text{news\_emb}} = \frac{1}{M}\sum_{i=1}^M \text{LM}(\text{news}_i), C_{\text{chart\_emb}} = \text{TS2Vec}(p_{t'}) 
\]

\[
Q_{\text{multimodal}} = \text{UMAP}(\bar Q_{\text{news\_emb}}) + Q_{\text{chart\_emb}}  
\]

\[
C_{\text{multimodal}} = \text{UMAP}(\bar C_{\text{news\_emb}}) + C_{\text{chart\_emb}} 
\]

\[
SIM(Q_{\text{multimodal}}, C_{\text{multimodal}}) = 1 - \frac{Q_{\text{multimodal}} \cdot C_{\text{multimodal}}}{||Q_{\text{multimodal}}|| ||C_{\text{multimodal}}||}
\]

\subsection{Random Sampling} 
Given some query timestep $t$, the random sampling samples from all $t'$ such that $t' < t$ randomly. We use random sampling to observe the differences in performance boost when using random sampling versus some other ranking method, thus verifying that our ranking methods do indeed catch patterns from the past that in turn help model BTC price movement. For each query, we randomly sample top 30 past patterns 100 times and calculate the performance of random sampling 100 times with these 100 sampling cases to get a better idea of how random sampling performs. 

\section{Experimental Results}

\subsection{Performance Comparison} 
We can make interesting observations from the results in Table 1.
\begin{itemize}
    \item On average all of our IR-based FE improves over the baseline of no FE, with the multimodal strategy performing the best. Using the top 5 most similar multimodal embeddings has the best F1 score of \textbf{0.628} and we will be using this model for backtesting in section 5.2. The TS2Vec and multimodal strategies may have improved performance had we used a longer $l$, but we leave this investigation for future research. 
    \item For our multimodal strategy, we separately calculated the accuracy of the model for the cases when it predicts $c_0$ or $c_1$ and when the ground truth action is also $c_0$ or $c_1$. Considering this case is important because when the model chooses $c_2$ it does nothing so it does not incur any profit or loss. If the model chooses $c_0$ or $c_1$ but the ground truth is $c_2$ then even if the model's decision is wrong, it would not result in a huge profit or loss since the volatility for that time frame would be small. Significant gains or losses happen when the model predicts $c_0$ or $c_1$ and when the ground truth action is also $c_0$ or $c_1$. In this case the multimodal strategy outperforms the no IR-based FE baseline by more than 5\% on average, while using the top 5 MultiModal IR-based FE outperforms no FE by close to 10\%.   
\end{itemize}

\subsection{Backtest on Test set}

\begin{figure}
    \centering
    \includegraphics[width=0.4\textwidth]{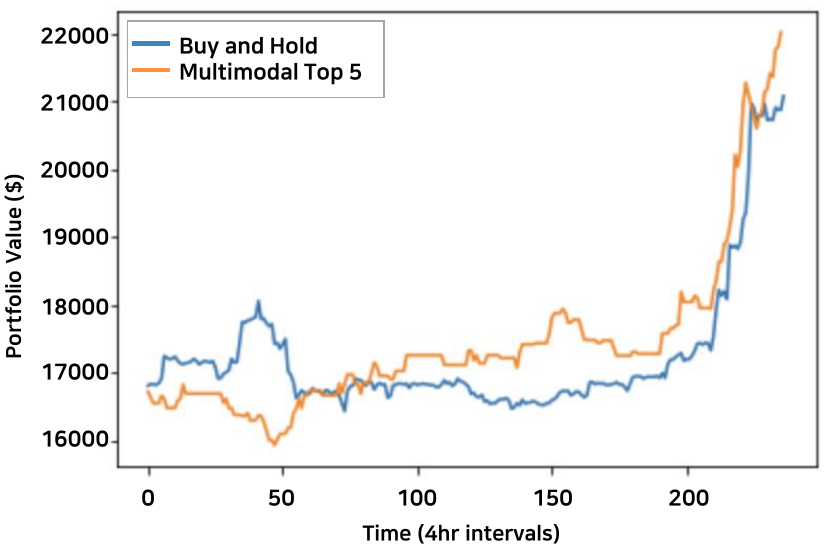}
    \caption{Backtest results.}
    \label{fig:backtest}
\end{figure}

The following assumptions were made for back testing: (1) We assume no take profit and a stop loss of 0.75\%. (2) We use commission rate of 0.04\%, equivalent to the maker fee when trading BTC/USDT perpetual in Binance. As a result in figure \ref{fig:backtest} shows, our model outperforms buy and hold for the duration of the test set. It is notable that the model predicts $c_1$ (short) effectively when the BTC prices are falling (e.g. around index 50), gaining edge over buy and hold. Also the model predicts $c_2$ (hold) very well (e.g. around index 100) for periods when there is less volatility and also predicts $c_0$ (long) around index 200 when the prices actually began to soar.

\section{Conclusion and Future Work}
In this research, we investigated how chart pattern matching can be incorporated into a BTC directional prediction model training framework. Among our proposed pattern matching methods, the multimodal embedding based method is the most effective. Overall, chart pattern matching based feature engineering seems promising and it can be further explored or coupled with other modeling techniques to more accurately forecast the volatile price movement of BTC. 

\section{Presentor Bio}
Minsuk Kim received his B.S. degree in mathematics from Stanford University, Stanford, CA, USA, in 2021. He is currently working as an AI Scientist and an Engineer at MindsLab in Korea. His research interests include financial machine learning and natural language processing. 

\section{Company Portrait}
We are primarily focuses on creating machine learning based trading bots and indicators for trading crypto assets.

\bibliographystyle{acm-reference-format}
\bibliography{acmart}
\end{document}